\documentclass[a4paper,11pt]{article}

\usepackage{jcappub} 

\usepackage{graphicx}
\usepackage{soul}
\usepackage{dcolumn}
\usepackage{color}
\usepackage[dvipsnames]{xcolor}
\usepackage{enumitem}
\usepackage{cancel}
\usepackage{mathtools}
\usepackage{hyperref}
\usepackage{orcidlink}
\usepackage{cases}

\newcommand{\tot}{{\rm tot}}
\newcommand{\pfrac}[2]{{\left(\frac{#1}{#2}\right)}}
\newcommand{\los}{{\rm los}}

\newcommand{\SMSM}{{\rm SS}}
\newcommand{\SD}{{\rm SD}}
\newcommand{\DD}{{\rm DD}}
\newcommand{\eff}{{\rm (eff)}}
\newcommand{\MAX}{{\rm (max)}}
\newcommand{\DM}{{\rm DM}}
\newcommand{\rel}{{\rm rel}}

\arxivnumber{2410.07324} 
\title{Constraints on Long-Ranged Interactions Between Dark Matter and the Standard Model}
\date{\today}

\author[a,b,c]{Zachary Bogorad\,\orcidlink{0000-0001-9913-6474}\,}
\emailAdd{zbogorad@fnal.gov}

\author[a,b]{Peter W.~Graham\,\orcidlink{0000-0002-1600-1601}\,}

\author[a,d]{Harikrishnan Ramani\,\orcidlink{0000-0003-3804-8598}\,}

\affiliation[a]{Stanford Institute for Theoretical Physics, Department of Physics, Stanford University, Stanford, CA 94305, USA}
\affiliation[b]{Kavli Institute for Particle Astrophysics \& Cosmology, Department of Physics, Stanford University, Stanford, CA 94305, USA}
\affiliation[c]{Fermi National Accelerator Laboratory, Batavia, IL 60510, USA}
\affiliation[d]{Department of Physics and Astronomy, University of Delaware, Newark, DE 19716, USA}

\abstract{Dark matter's existence is known thanks to its gravitational interaction with Standard Model particles, but it remains unknown whether this is the only force present between them. While many searches for such new interactions with dark matter focus on short-range, contact-like interactions, it is also possible that there exist weak, long-ranged forces between dark matter and the Standard Model. In this work, we present two types of constraints on such new interactions. First, we consider constraints arising from the fact that such a force would also induce long range interactions between Standard Model particles themselves, as well as between dark matter particles themselves. Combining the constraints on these individual forces generally sets the strongest constraints available on new Standard Model-dark matter interactions. Second, we consider the possibility of constraining new long-ranged interactions between dark matter and the Standard Model using the effects of dynamical friction in ultrafaint dwarf galaxies, especially Segue I. Such new interactions would accelerate the transfer of kinetic energy from stars to their surrounding dark matter, slowly reducing their orbits; the present-day stellar half-light radius of Segue I therefore allows us to exclude new forces which would have reduced stars' orbital radii below this scale by now.}

\begin{document}
\maketitle
\flushbottom




\section{Introduction}\label{sec:Introduction}

The gravitational effects of dark matter have now been detected in a variety of systems, leaving the ubiquity of dark matter in the universe essentially beyond dispute. Non-gravitational signatures of dark matter remain elusive, however, and in fact it remains unknown whether dark matter interacts with the Standard Model through any mediator except gravity at all. The existence and nature of such additional interactions are thus crucial assumptions to almost all attempts to establish the identity and properties of dark matter.

New interactions between dark matter and the Standard Model can take a variety of forms \cite{DMPortalsReview}, with different dark matter models and forces constrained by different experimental and astrophysical observations. At low energies, however, most such interactions reduce to one of a few possible classical potentials \cite{MoodyWilczekNewForces} (see e.g. Ref. \cite{WarpedDarkSector} for an exception), with the nature of the mediator determining parameters such as the potential's range and species-dependent couplings. 

Most of the existing work on new dark matter-Standard Model interactions has focused on short-range, contact-like forces (e.g. the widely studied WIMP paradigm \cite{WIMPReview}); conversely, in this work, we focus on forces that have long (but generally finite) ranges. We will focus in particular on roughly mpc- to kpc-scale forces, although most of this work would apply to longer-range forces as well, and some can be extended to somewhat shorter ranges.

Such long-range forces can have major implications for our understanding of astrophysics. In fact, even short-range interactions of dark matter particles can have major implications for dark matter structures, as studies of self-interacting dark matter have shown \cite{SIDM1, SIDM2, SIDM3, SIDM4}. Long-range forces, however, can have much larger effects, thanks to the potential for collective effects of many dark matter particles causing or experiencing a force coherently (see, for example, Ref. \cite{CosmicAxionForce}). As a result, long-range forces---both between dark matter particles alone, and between dark matter particles and the Standard Model---can have far more dramatic consequences for astrophysical structures. In particular, our understanding of the distribution of dark matter throughout the universe---whether in the CMB or within galaxies and galaxy cluster---is based on its interactions with the Standard Model, assuming that the only such interaction is gravity. Additional long-range interactions of dark matter and the Standard Model could therefore dramatically modify the dark matter densities that we infer. In this work, we study both this effect and other implications of such a new force.

In most of this work, we focus for concreteness on forces described by Yukawa potentials, which are a highly general prediction for spin-independent forces. We will parametrize such potentials $V$ as
\begin{align}
    V &= q_{\rm S}q_{\rm D}\alpha_{\rm SD} G\frac{m_{\rm S}m_{\rm D}}{r}e^{-r/\lambda} \label{eq:YukawaSD}
\end{align}
where $\alpha_{\rm SD}$ is the characteristic strength of the new force compared to gravity, $G$ is Newton's constant, $m_{\rm S}$ ($m_{\rm D}$) is the mass of the Standard Model (dark matter) particle, and $q_{\rm S}$ ($q_{\rm D}$) is the charge of that particle, which we will assume to be order-one but potentially with either sign. Note that our parametrization of the strength relative to gravity means that we will frequently consider $\alpha_{\rm SD} \gg 1$ in this work. Such couplings will nonetheless generally be perturbative; we discuss exceptions in Appendix \ref{app:FiniteRangeHeating}.

Crucially, a new Standard Model-dark matter force, as in Eq. \eqref{eq:YukawaSD}, involves a mediator coupled to both Standard Model particles and dark matter particles, and therefore also leads to a new force between Standard Model particles alone and between dark matter particles alone. These forces need not have similar strengths: letting $g_S$ ($g_D$) be the coupling to a Standard Model (dark matter) particle of charge $q_S=1$ ($q_D=1$), we have two independent degrees of freedom setting the three characteristic strengths: $\alpha_\SMSM = g_S^2/(4\pi)$, $\alpha_\DD = g_D^2/(4\pi)$, and $\alpha_\SD = g_Sg_D/(4\pi)$, with $\alpha_\SMSM$ ($\alpha_\DD$) defined analogously to $\alpha_\SD$ but for Standard Model-Standard Model (dark matter-dark matter) forces 

One additional assumption is required for the dynamical friction results of this work: stars must carry significant net charges under the considered new force. Our dynamical friction constraints will thus apply to, for example, new forces gauging baryon minus lepton number ($U(1)_{\rm B-L}$), but would not apply to dark matter coupled through a kinetically-mixed dark photon. Note that there is no analogous requirement for the dark matter, which may or may not have a net-neutral mixture of charges.

The presence of such new interactions between stars and dark matter accelerates the process of dynamical friction: the gradual transfer of energy from stars to the cold dark matter around them, whether through gravity or through a new long-ranged force. In particular, the presence of new forces stronger than gravity can cause stars to give up order-one fractions of their momentum to the surrounding dark matter over cosmological timescales, causing them to de-orbit towards the centers of their containing galaxies. The goal of this work will therefore be to use the observation of stellar distributions today---especially in ultrafaint dwarf galaxies (UFDs), where this effect is relatively large---to set constraints on new forces between stars and dark matter.

The remainder of this work is organized as follows: In Section \ref{sec:CombinedConstraints}, we present constraints on new long-ranged interactions between dark matter and the Standard Model from combining constraints on dark matter-only and Standard Model-only forces, since any force between the two must necessarily couple to both. We therefore briefly review these individual constraints and present the resulting constraints on $\alpha_\SD$. We then describe a new constraint directly on $\alpha_\SD$ based on dynamical friction in UFDs in Section \ref{sec:DFinUFDs}. We begin by summarizing the observation status of UFDs and the effects of dynamical friction itself on the evolution of stellar distributions within them. We then consider a couple of related subtleties, including the modification of UFD dark matter distributions by the presence of a new force and the influence of plasma effects on our results. Finally, in Section \ref{sec:Conclusion}, we summarize the observational status of long-ranged forces between dark matter and the Standard Model.

\section{Combined Constraints}\label{sec:CombinedConstraints}

One source of constraints on long-range interactions between dark matter and the Standard Model arises from combining separate constraints on the dark matter coupling and the Standard Model coupling to the new force mediator, i.e. by using \cite{GeometricMeanConstraint}
\begin{align}
    \alpha_{\rm SD} = \sqrt{\alpha_{\rm SS}\alpha_{\rm DD}}; \label{eq:GeometricMean}
\end{align}
see the discussion below Eq.~\eqref{eq:YukawaSD}. We will refer to these as ``combined constraints'' for convenience. For many classes of new forces, these will set the strongest constraints in this work, thanks in large part to the extremely strong limits which can be placed on new long-range interactions of Standard Model particles. We therefore begin by reviewing the present state of constraints on $\alpha_\SMSM$ and $\alpha_\DD$, before turning to limits set directly on dark matter-Standard Model interactions in the next section.

We recently discussed constraints on long-range dark matter self-interactions in Ref.~\cite{LRDMSI}. For attractive self-interactions, the strongest constraints at long-but-finite ranges currently arise from a combination of observations of the shockwaves observed in the gas of the Bullet Cluster and the spatial coincidence of the dark matter and stars in that cluster. Repulsive self-interactions are currently better constrained using the binding of ultrafaint dwarf galaxy's halos instead. As we noted in Ref.~\cite{LRDMSI}, it may eventually be possible to set stronger constraints on both types of self-interactions based on their effects on structure formation, but this analysis has not been performed to-date.

One additional constraint on $\alpha_\DD$ has been published more recently: Ref.~\cite{EFTofLSSaDD} presents constraints on attractive dark matter-dark matter interactions with ranges of $\sim 40$ kpc or more based on their effects on large-scale structure. In particular, forces with ranges less than the inverse Hubble constant suppress the matter power spectrum and introduce features around the force mediator's Jeans scale; this can be compared against results from galaxy surveys in order to constraint such forces.

Finally, we note that---in the case of forces where both signs of charge are present within dark matter, such that the dark matter forms a plasma---one additional effect may lead to significantly stronger constraints in the foreseeable future: plasma instabilities. In particular, Ref.~\cite{DMPlasmaInsabilities} shows that collisional instabilities would lead to exponential growth of perturbations in various astrophysical systems for many new force parameters of interest. While this has not been used to set constraints to-date, it may be possible to do so in the future; see Section \ref{sub:PlasmaEffects} for further discussion.

New long-ranged forces acting on the Standard Model alone have been constrained with a variety of experiments and observations, although their interpretation is complicated by the question of how similar they are to gravity. In particular, long-ranged forces are generally easier to constrain when they lead to larger apparent violations of the equivalence principle. This is because equivalence principle-violating effects can be detected via differential accelerations between different materials, whereas equivalence principle-preserving ones can be detected only through the difference between the $r^{-1}$ behavior of gravity and the Yukawa potential of Eq.~\eqref{eq:YukawaSD}, which is only weakly suppressed on length scales $r < \lambda$. 

The extent to which new forces violate the equivalence principle can vary dramatically even between simple models: forces coupled to baryon number (e.g.~from gauging $U(1)_B$) are quite similar to gravity in many contexts since the masses of atoms are nearly proportional to their atomic numbers, whereas forces coupled to baryon minus lepton number (e.g. from gauging $U(1)_{B-L}$) are far more equivalence principle-violating since the neutron fraction of atomic nuclei varies considerably. However, even baryon number-coupled forces show equivalence principle violation of order 1\% due to nuclear binding energies, and it is generally challenging for models to be much more gravity-like than this.

A variety of models with new gravity-like forces have been proposed, including ones based on scalar gravity, bigravity, dilatons, etc. We will not attempt to review these models, and the extent to which the resulting forces can be equivalence principle-conserving, in this work; see, for example, Refs. \cite{ScalarGravity, EffectiveEPV1, EffectiveEPV2, ScalarDarkForces}. We will instead present constraints based on the two limiting cases of a new force's equivalence principle violation: order-one violation (i.e., in the language of Eq.~\eqref{eq:YukawaSD}, a force which couples with $q_S \sim 1$ for one test mass but $q_S \ll 1$ for the other) and zero violation (i.e.~a force exactly proportional to any Standard Model test object's total mass, with $q_S=1$ always). Typical models of new forces will then lie somewhere between these extremes, and the resulting constraints can generally be obtained straightforwardly by interpolating between these limits.

The strongest limit on equivalence principle-violating forces at the long ranges we consider generally comes from the MICROSCOPE experiment, which looks for differential acceleration between two concentric cylinders with different atomic compositions in orbit around the Earth \cite{MICROSCOPE, MICROSCOPEDilaton}. The lack of an observed difference between these accelerations constrains new forces with order-one equivalence principle violation to strengths
\begin{align}
    \left|\alpha_{\rm SS}^{\rm (max.~EPV)}\right| \lesssim {\rm few}\times 10^{-15}, \label{eq:MICROSCOPELimit}
\end{align}
independent of the new force's range for the ranges we consider, which are all much larger than the cylinders' distance from Earth's surface ($\sim 700$ km). As we noted above, many models of new forces lead to equivalence principle violation far below maximal; in this case, the limit of Eq.~\eqref{eq:MICROSCOPELimit} should be weakened in proportion to the fractional acceleration difference expected between titanium and platinum for that model.

Constraining long-range new forces that satisfy the equivalence principle is substantially harder, since such forces become indistinguishable from being a component of gravity in the limit that the force range is much larger than the distance over which the force is tested. Certainly such new forces cannot be stronger than gravity at these lengthscales\footnote{In principle, this might be avoidable in a fine-tuned scenario in which a finite-range repulsive new force with strength $|\alpha|$ almost exactly cancels infinite-range attractive gravity strengthened relative to our naive prediction by a factor of $|\alpha|+1$, leading to an apparent force of gravity matching observation. (Note that such a force would also need to e.g. reproduce the bending of light predicted by general relativity.) Such a scenario appears highly unmotivated, however, so we will not consider it further.}:
\begin{align}
    \left|\alpha_{\rm SS}^{\rm (no~EPV)}\right| \leq 1 \qquad (\lambda \gg 100~\mu{\rm m}).
\end{align}
Otherwise, however, the only means of detecting such a force on length scales shorter than the force range (i.e. for $r \ll \lambda$) is through the small modification of the shape of the total ``gravitational'' potential, since
\begin{align}\begin{split}
    V_{\rm new} &= \alpha_{\rm SS} G\frac{m_{1}m_{2}}{r}e^{-r/\lambda} \\
    &\approx \alpha_{\rm SS} G m_{1}m_{2} \left(\frac{1}{r} - \frac{1}{\lambda} + \frac{r}{2\lambda^2} + ...\right).
\end{split}\end{align}
Note that the middle term in the second line is an unobservable distance-independent contribution to the energy, and thus the leading effect of replacing a fraction $\alpha_{\rm SS}$ of gravity with this long-range equivalence principle-preserving new force is a change in the distance-dependence of the force of gravity suppressed by $(r/\lambda)^2$. The strongest constraints that can be placed at the ranges of interest in this work then arise from the modification of planetary orbits that such a new force would induce, giving \cite{AlphaSSnEPVPlanets, AlphaSSBook, AlphaSSReview}
\begin{align}
    \left|\alpha_{\rm SS}^{\rm (no~EPV)}\right| \lesssim 7\times10^{-10} \left(\frac{\lambda}{\rm AU}\right)^2 \qquad (\lambda \gg {\rm AU}).
\end{align}

\begin{figure}
    \centering
    \includegraphics[width=0.7\linewidth]{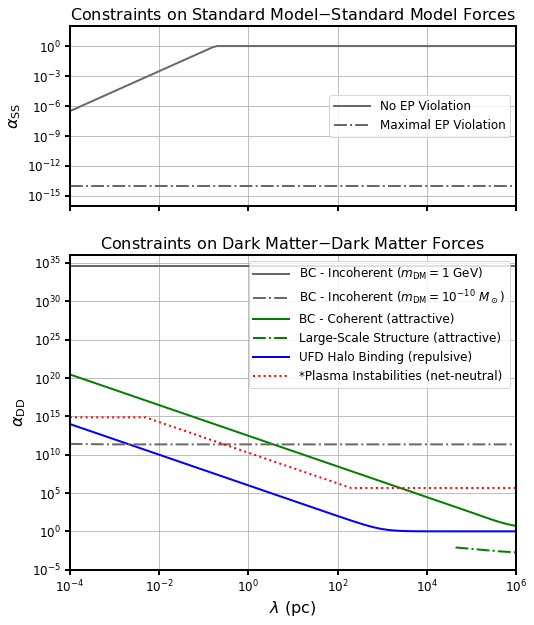}
    \caption{Existing constraints on new long-range forces on Standard Model particles (top) and on dark matter particles (bottom), i.e. on $\alpha_\SMSM$ and on $\alpha_\DD$; see Section \ref{sec:CombinedConstraints} for discussion. \textit{Top:} Shown are constraints on SM-SM forces in both the maximal and zero equivalence principle violation limits. In the presence of non-negligible equivalence principle violation, the strongest constraint at all ranges comes from the MICROSCOPE experiment \cite{MICROSCOPE}. In the absence of any EP violation, the constraints arise from planetary observations \cite{AlphaSSnEPVPlanets, AlphaSSBook, AlphaSSReview} or, once the range is too large to distinguish the force from gravity on observed length scales, from the known strength of gravity. \textit{Bottom:} A summary of constraints on a new, long-range, DM-DM force.   One constraint shown arises from incoherent scattering in the Bullet Cluster \cite{BCFirstDMSILimits, LDMReviewKLZ, BCDMSILimits2, BCDMSILimits3} (where we have shown two example dark matter masses) which is applicable for attractive, repulsive, and net-neutral dark matter.  Also shown is a constraint from coherent interactions in the bullet cluster which is applicable for attractive forces \cite{LRDMSI}. An additional constraint shown is from effects of attractive forces on the formation of large-scale structure \cite{EFTofLSSaDD}.  A final existing constraint is from self-binding in the presence of UFDs and applies to repulsive forces only \cite{LRDMSI}. Also shown is the potential constraint that could be obtained in the future from excluding exponentially growing plasma instabilities in astrophysical systems, in the case of net-neutral dark matter  \cite{DMPlasmaInsabilities}.}
    \label{fig:IndividualConstraints}
\end{figure}

A summary of these existing constraints on long-range dark matter-dark matter and Standard Model-Standard Model forces is plotted in Fig.~\ref{fig:IndividualConstraints}. The constraints on $\alpha_{\rm SD}$ which result from combining these constraints are included in Figs.~\ref{fig:AttractiveNEPV}--\ref{fig:NeutralWEPV}, where we compare them to the other constraints discussed in this work.

\section{Dynamical Friction in Ultrafaint Dwarf Galaxies}\label{sec:DFinUFDs}

We now turn to deriving a new constraint on long range interactions between dark matter and the Standard Model from the enhanced dynamical friction that would arise in the presence of such a force, which would decelerate the stars within dark matter halos. In particular, the stars in ultrafaint dwarf galaxies would gradually de-orbit towards the centers of their halos, reducing the half-light radius of such galaxies over time. Conservative assumptions about where the stars began---namely that they formed within their halo's scale radius---combined with observations of these galaxies' present-day half-light radii then lead to a maximum allowed new force coupling, beyond which the stars would have necessarily de-orbited to radii less than those observed.

For sake of comparison, it is worth noting the strongest existing constraint directly on $\alpha_\SD$: Ref. \cite{AlphaSDDirectTorsion} sets a constraint on equivalence-principle violating forces on a torsion balance which are directed towards the Milky Way center. This constrains
\begin{align}
    \alpha_{\rm SD}^{\rm (max.~EPV)} \lesssim 10^{-4} \qquad (\lambda \gg 8~{\rm kpc})
\end{align}
with $8$ kpc the distance to the Milky Way center. This can likely be extended to shorter ranges, albeit suppressed by a factor of $(8~{\rm kpc}/\lambda)^2$, based on the effects of dark matter gradients; see Ref. \cite{LRDMSI}. We do not consider this in detail, however, since this constraint is subleading compared to the combined constraints discussed in Section \ref{sec:CombinedConstraints}. Some other constraints on $\alpha_\SD$ are presented in Ref. \cite{GeometricMeanConstraint, SupernovaConstraints}, but we will not review them here since they are generally subdominant and carry some additional model dependence.

\subsection{Properties of Ultrafaint Dwarfs}\label{sub:UFDProperties}

We begin by briefly reviewing the properties of ultrafaint dwarf galaxies; for more detailed discussion of UFDs see, for example, Refs. \cite{UFDData1, UFDData2}. Several dozen UFDs have now been observed, and many others are likely to be found in the near future. UFDs are distinguished by simultaneously having both some of the smallest virial masses (expected to be $\sim 10^9~M_\odot$, although only the central densities of UFD halos can actually be observed) known for dark matter-dominated systems, and some of the highest dark matter mass fractions of such systems. They are also generally quite old (as we know, in large part, thanks to their low metallicities). The extremely low luminosities of these objects make detailed observations difficult, however, so their properties often carry large uncertainties.

The UFD that we will focus on here is Segue 1, whose properties are comparatively well-known \cite{UFDData2, Segue1Data1, Segue1Data2, Segue1Data3}. We note, however, that our analysis can be straightforwardly extended to many other UFDs as well. Segue 1 is a UFD located approximately $23$ kpc from the Sun, with a luminosity of roughly $300~L_\odot$ (and thus $\sim 300$ stars), a half-light radius of approximately $32.3$ pc, and a stellar line-of-sight velocity dispersion of $3.7$ km s$^{-1}$. Simulations of UFD population evolution predict that Segue I likely has a mass of approximately $10^9~M_\odot$, while estimates of its age, based largely on the metallicity of its stars, suggest an age of order 10 Gyr \cite{SegueOrigin1, SegueOrigin2, SegueOrigin3}. We will adopt these values for our calculations (up to corrections discussed in Section \ref{sub:UFDChanges}), although we note that their uncertainties vary from several percent to order-one. None of these uncertainties should affect our conclusions by more than order-one factors, however.

From these observed quantities, we can now derive several additional properties of Segue I that will be important for our calculations below: the size of its dark matter halo, the temperature (or, equivalently, velocity) distributions of the stars and the dark matter, and the potential energy of the galaxy as a function of the stellar half-light radius (since we will be interested in scenarios in which this radius has changed over time). These properties depend, however, on the profile assumed for the UFD's dark matter, which is not precisely known. We therefore perform our calculations with three separate profiles, in order to illustrate our result's robustness against uncertainty in the dark matter halo shape: Dehnen spheres with both $\gamma=0$ and $\gamma=1$ \cite{DehnenSpheres1, DehnenSpheres2},
\begin{align}
    \rho_\chi(r) = \frac{(3-\gamma)M_{\chi,\tot}}{4\pi R_\chi^3}\pfrac{r}{R_\chi}^{-\gamma}\left(1+\frac{r}{R_\chi}\right)^{\gamma-4}
\end{align}
and an NFW profile with a concentration of 15,
\begin{align}
    \rho_\chi(r) = \frac{M_{\chi,\tot}}{4\pi R_\chi^3 \mathcal{N}}\pfrac{r}{R_\chi}^{-1}\left(1+\frac{r}{R_\chi}\right)^{-2}
\end{align}
with $M_{\chi,\tot} \sim 10^9\,M_\odot$ the total mass of dark matter in Segue I (or within 15 scale radii, for the NFW profile), $R_\chi$ the dark matter's scale radius (which we estimate below), and $\mathcal{N} \sim 1.8$ a normalization factor. We assume the stars in Segue I are distributed in a Plummer sphere \cite{PlummerSpheres1, PlummerSpheres2},
\begin{align}
    \rho_*(r) = \frac{3M_{*,\tot}}{4\pi R_*^3}\left(1+\frac{r^2}{R_*^2}\right)^{-5/2},
\end{align}
with $M_{*,\tot} \sim 300 M_\odot$ the total stellar mass in Segue I and $R_*$ the stellar scale radius (related to the half-light radius $R_{\frac{1}{2}}$ by $R_* = \sqrt{2^{2/3}-1}R_{\frac{1}{2}}$).

These profiles should be truncated at some radius due to the effects of tidal forces \cite{TidalTruncation}. While the tidal radius of Segue I is not known precisely, it is expected to be at least of order $R_\chi$, given the fairly small eccentricity of Segue I's orbit \cite{SegueOrigin2}. This truncation is therefore negligible for our purposes, since it affects only dark matter much farther from the stars than the largest impact parameters we will consider (see Section \ref{sub:DFEffects} and Appendix \ref{app:FiniteRangeHeating}).

In this section, we consider the effects of gravity alone, without any new forces; we account for the effects of the new forces of interest in this work in Section \ref{sub:UFDChanges}.

The dark matter scale radius $R_\chi$ can be estimated from our knowledge of the stellar velocity dispersion and half-light radius. In particular, the stellar line-of-sight velocity dispersion is approximately \cite{SigmaLOSRelation}
\begin{align}
    \left\langle\sigma_{*,\los}^2\right\rangle \approx \frac{GM_{\frac{1}{2}}}{3R_{\frac{1}{2}}}
\end{align}
with $M_{\frac{1}{2}}$ the total mass within a half-light radius of the UFD's center. Requiring this to match the observed value of $\sqrt{\langle\sigma_{*,\los}^2\rangle} = 3.7{\rm~km~s}^{-1}$ gives dark matter scale radii of 532 pc (for $\gamma=0$), 1859 pc (for $\gamma=1$), or 975 pc (for NFW).

We now turn to estimating the temperatures (or velocity dispersions) of the stars and the dark matter. These can be obtained by invoking hydrostatic equilibrium \cite{HydrostaticEquilibriumGalaxies}, such that
\begin{align}
    \frac{dP_{*}}{dr} = \frac{d}{dr}\pfrac{T_*(r)\rho_*(r)}{m_*} = -\frac{G\rho_*(r)}{r^2}M_\tot(r) \label{eq:HydrostaticEquilibriumS}
\end{align}
with $P_*(r)$ the pressure of the collection of stars at radius $r$, $T_*(r)$ its temperature (in the stellar velocity sense), $\rho_*(r)$ the stellar density, $m_*$ the typical mass of stars in Segue I (which we take to be $1~M_\odot$, since this is fairly typical; we are not aware of any more precise measurement of it) and $M_{\rm tot}$ the total mass of stars and dark matter within the radius $r$. An identical expression, but with the starred quantities replaced by the $\chi$ quantities, must also hold for dark matter:
\begin{align}
    \frac{dP_{\chi}}{dr} = \frac{d}{dr}\pfrac{T_\chi(r)\rho_\chi(r)}{m_*} = -\frac{G\rho_\chi(r)}{r^2}M_\tot(r) \label{eq:HydrostaticEquilibriumD}
\end{align}
The general solutions of these equations are messy and uninsightful. We, however, will use only the velocity dispersions at the stellar scale radius, since this is where the majority of the heat transfer we eventually wish to calculate will take place: smaller radii include little volume, while larger radii have low stellar densities. The velocity dispersions at the stellar scale radius can be approximated by, for the $\gamma=0$ case,
\begin{subequations}\begin{align}
    \sigma_*(R_*) &= \sqrt{\frac{2GM_{\chi,\tot}R_*^2}{3R_\chi^3}} \label{eq:SigmaSRS} \\
    \sigma_\chi(R_*) &= \sqrt{\frac{GM_{\chi,\tot}}{30R_\chi}} \label{eq:SigmaDRS}
\end{align}\end{subequations}
where we have included only the contributions from dark matter since the leading-order stellar contributions are suppressed by $\rho_*/\rho_\chi \ll 1$. The analogous expressions for $\gamma=1$ and NFW profiles are given in Appendix \ref{app:AnalyticExpressions}.

Finally, we compute the potential of Segue I arising from its self-gravitation, as a function of the stellar half-light radius. We omit the contribution from the self-gravitation of the dark matter alone, since the negligible total mass of stellar matter in Segue I compared to the total mass of dark matter means that the dark matter distribution is essentially constant in time. (This is not an entirely rigorous simplification, but no superior alternatives are known and, numerically, it has been found to work well; see Appendix A of Ref.~\cite{DFofStarsinUFDs}.) We can then break the potential into three terms: the self-gravitation of the stars $U_{**}$, the potential given to stars by dark matter $U_{*\chi}$, and the potential given to dark matter by stars $U_{\chi*}$. (The latter two are not redundant, as we consider only the potential of e.g.~dark matter at radii $r<r'$ when evaluating the potential of a star at radius $r'$.) For the $\gamma=0$ profile, these evaluate to
\begin{subequations}\begin{align}
    U_{**} &= \frac{3\pi GM_{*,\tot}^2}{32R_*} \label{eq:USS0} \\
    U_{*\chi} &= \frac{4GM_{*,\tot}M_{\chi,\tot}R_\chi^2}{\pi^{3/2}R_*^3}G^{3,3}_{3,3}\left(\frac{R_*^2}{R_\chi^2}\left|\begin{matrix}1 & \frac{3}{2} & 2 \\[2pt] \frac{5}{2} & \frac{5}{2} & 3\end{matrix}\right.\right) \label{eq:USD0} \\
    U_{\chi*} &= \frac{4GM_{*,\tot}M_{\chi,\tot}R_\chi^2}{\pi^{3/2}R_*^3}G^{3,3}_{3,3}\left(\frac{R_*^2}{R_\chi^2}\left|\begin{matrix}1 & 1 & \frac{3}{2} \\[2pt] \frac{3}{2} & \frac{5}{2} & 3\end{matrix}\right.\right) \label{eq:UDS0}
\end{align}\end{subequations}
where $G^{\cdot,\cdot}_{\cdot,\cdot}\left(\cdot\left|\begin{matrix}\cdot & ... & \cdot \\ \cdot & ... & \cdot\end{matrix}\right.\right)$ is the Meijer G-function. We again defer the expressions for NFW and $\gamma=1$ profiles to Appendix \ref{app:AnalyticExpressions}.

\subsection{Effects of Dynamical Friction}\label{sub:DFEffects}

Dynamical friction is the way in which the components of galaxies thermalize, with the average kinetic energies of stars (as a whole; not their component particles) and dark matter particles gradually equalizing. For dark matter models with particles much lighter than a solar mass, this corresponds to a transfer of kinetic energy from stars to the dark matter around them. In this section, we calculate the rate of dynamical friction for gravity alone; the corresponding calculation for Yukawa forces, which differs only in small ways for most of the relevant parameter space, is presented in Appendix \ref{app:FiniteRangeHeating}.

For Coulomb-like scattering with a velocity-dependent momentum-transfer cross-section $\sigma_T=\sigma_0 v^{-4}$ and $m_* \gg m_\chi$, the rate of heat transfer from stars to dark matter is given by \cite{CoulombHeatingRate1, CoulombHeatingRate2, CoulombHeatingRate3}
\begin{align}
    \frac{d^2U_\chi}{dt\,dV} &= \frac{2\rho_*\rho_\chi\sigma_0(T_*-T_\chi)}{\sqrt{2\pi}m_*^2v_{\rm th}^3} \label{eq:dUdtdV}
\end{align}
with $dU_\chi/dV$ the energy per unit volume of the dark matter, $T_*$ ($T_\chi$) the temperature of the stars (dark matter), $m_*$ ($m_\chi$) the mass of the stars (dark matter), and
\begin{align}
    v_{\rm th} = \sqrt{\frac{T_*}{m_*}+\frac{T_\chi}{m_\chi}}.
\end{align}
We emphasize again that $T_*$ is the temperature corresponding to the distribution of stellar velocities, and not the temperature of their interiors.
 
In the case of gravity, which is infinite-ranged, we can rewrite this heating rate as \cite{DFofStarsinUFDs, GalacticDynamics}
\begin{align}
    \frac{d^2U_\chi}{dt\,dV} &= 4\sqrt{2\pi}G^2\rho_*\rho_\chi\frac{m_*\sigma_*^2-m_\chi\sigma_\chi^2}{(\sigma_*^2+\sigma_\chi^2)^{3/2}}\ln\Lambda_G
\end{align}
where $\sigma_*$ ($\sigma_\chi$) is the velocity dispersion of the stars (dark matter), such that $T_{*,\chi}=m_{*,\chi}\sigma_{*,\chi}^2$, and $\ln\Lambda_G$ is a factor arising from regulating the forward divergence of the Coulomb cross-section; see Appendix \ref{app:FiniteRangeHeating} for the analogous calculation for finite-range forces. Specifically \cite{DFofStarsinUFDs},
\begin{align}
    \Lambda_G^2 &= \frac{G^2m_*^2 + b_{\rm max}^2v_{\rm th}^4}{G^2m_*^2 + b_{\rm min}^2v_{\rm th}^4} \label{eq:CoulombLog}
\end{align}
with $b_{\rm max}$ ($b_{\rm min}$) the maximum (minimum) impact parameter. These are set by
\begin{align}
    b_{\rm max} &= R_* \\
    b_{\rm min} &= \frac{Gm_*}{\sigma_*^2+\sigma_\chi^2}
\end{align}
where $R_*$ is the stellar scale radius for the galaxy. The former is thus an approximate upper bound for how far away stars can be from $\chi$ particles when transferring energy to them, while the latter is the $90^\circ$ deflection radius below which smaller impact parameters cease to increase momentum transfer. (We conservatively limit $b_{\rm max}$ to $R_*$, as in Ref. \cite{GalacticDynamics}, rather than the $R_*\sigma_\chi/\sigma_*$ of Ref. \cite{DFofStarsinUFDs}, but this has little effect on our results.)

Computing the total heat transfer rate from stars to dark matter would require us to integrate Eq. \eqref{eq:dUdtdV} over the UFD's volume. While this could be performed numerically, we can obtain approximate analytic results for this total heating rate by taking advantage of the small variation in the temperatures of the stars and dark matter at radii $r \lesssim R_*$, where the majority of heat transfer happens; while not exact, this is sufficient for the order-one estimates in this work, and is numerically less expensive than the numerical alternative. In this approximation, the total heat transfer becomes
\begin{align}\begin{split}
    \frac{dU_\chi}{dt} \approx\,& 4\sqrt{2\pi}G^2 \frac{M_{*,\tot}M_{\chi,\tot}}{R_*^3}\mathcal{F} \frac{m_*\sigma_*(R_*)^2-m_\chi\sigma_\chi(R_*)^2}{(\sigma_*(R_*)^2+\sigma_\chi(R_*)^2)^{3/2}}\ln\Lambda_G(R_*) \label{eq:TotalHeatingRate}
\end{split}\end{align}
where $\mathcal{F}$ is a dimensionless form factor for the density overlap between the stars and the dark matter:
\begin{align}
    \mathcal{F} &= \frac{R_*^3\int dV \rho_*\rho_\chi}{M_{*,\tot}M_{\chi,\tot}}. \label{eq:FormFactorDef}
\end{align}
This form factor can be analytically evaluated for each of the halo profiles we consider in this work. For $\gamma=0$, this gives
\begin{align}
    \mathcal{F} &= \frac{2}{\pi^{5/2}}G^{3,3}_{3,3}\left(\frac{R_*^2}{R_\chi^2}\left|\begin{matrix}0 & \frac{1}{2} & 1 \\[2pt] \frac{3}{2} & 2 & \frac{5}{2}\end{matrix}\right.\right)
\end{align}
We again defer the NFW and $\gamma=1$ results to Appendix \ref{app:AnalyticExpressions}.

\begin{figure}[b]
    \centering
    \includegraphics[width=0.7\linewidth]{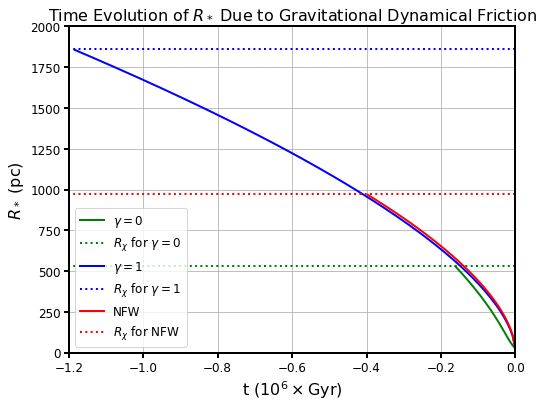}
    \caption{The stellar scale radius of Segue I at various times in the past, assuming dynamical friction from gravity alone and three different halo profiles: Dehnen spheres with $\gamma=0$ and $\gamma=1$ and an NFW profile. Also shown are the (constant) dark matter scale radii of each profile, obtained by combining the simulated total mass and the observed stellar velocity dispersion. The maximum times in the past for each line are chosen so that the stellar scale radius would have been greater than the dark matter scale radius, since the unphysical nature of such a galaxy is the basis for the constraints we set (see Sections \ref{sub:UFDChanges} and \ref{sub:PlasmaEffects}); while the times required for this are unphysical in the gravity-only case, the introduction of new forces is largely equivalent to rescaling the time, leading to significant evolution over physically-meaningful timescales.}
    \label{fig:TimeEvolutionDF}
\end{figure}

We can now translate this rate of energy loss by the stars into a rate of change for the stellar scale radius using our results from the previous section:
\begin{align}
    \frac{dR_*}{dt} = -\frac{dU_\chi/dt}{d(U_{**}+U_{*\chi}+U_{\chi*})/dR_*}.
\end{align}
This finally, must be integrated numerically in order to evaluate the time evolution of the stellar scale radius. The results of integrating up to $\sim1.2\times10^6$ Gyr backwards in time, including only the effects of gravity, are presented in Fig. \ref{fig:TimeEvolutionDF}. (We terminate the $\gamma=0$ and NFW integration earlier since we will not be interested in the behavior for $R_* > R_\chi$.) While this is a substantially longer time than the age of Segue I (or of the universe), we will see below that the presence of new forces accelerates dynamical friction essentially by a rescaling of time, such that an unphysical time of $\sim10^6$ Gyr may correspond to a physically meaningful time in the presence of, for example, a new force with $\alpha_\SD \sim 10^3$. The similarity of the three curves in Fig. \ref{fig:TimeEvolutionDF} illustrates our modest dependence on the assumed halo profile shape (although this model-dependence will be enhanced somewhat by the effects discussed in the next two sections).

In the presence of new forces, Fig. \ref{fig:TimeEvolutionDF} is modified for two reasons: First, there is a direct $\alpha_\SD^2$ dependence in the total heating rate (i.e. $G^2 \to \alpha_\SD^2 G^2$ in Eq. \eqref{eq:TotalHeatingRate}), which accelerates dynamical friction by a factor of $\alpha_\SD^2$. (There is additionally a logarithmic dependence on $\alpha_\SD$ and on the new force range $\lambda$ through a modified logarithm (see Eqs. \eqref{eq:YukawaCutoff} and \eqref{eq:YukawaLog} in Appendix \ref{app:FiniteRangeHeating}), but this is only significant near the shortest ranges where our bound applies). If this quadratic contribution were the only modification to the heating rate, we could immediately set constraints on new forces based on a conservative assumption about galaxy formation: that the initial stellar scale radius be smaller than the dark matter scale radius, since this is where stars are expected to form. The maximum allowed $\alpha_\SD$ (for $\lambda > R_*$) would then be given by
\begin{align}
    \alpha_{\SD}^\text{(na\"{i}ve max)} \sim \sqrt{\frac{-\tau_{\rm eq}^G}{10~{\rm Gyr}}}
\end{align}
with $\tau_{\rm eq}^G$ the time at which the scale radii would have been equal for the assumed profile with gravity alone (e.g. $\tau_{\rm eq}^G \sim -2.5\times10^6~{\rm Gyr}$ for $\gamma=0$).

However, there is a second modification to the rate of dynamical friction caused by the introduction of a new dark matter-Standard Model force. A new force of this type also implies the existence of dark matter-dark matter and Standard Model-Standard Model forces, as discussed in Section \ref{sec:CombinedConstraints}. While the small mass of Standard Model particles in UFDs makes the latter irrelevant, the introduction of a strong ($\alpha_{\rm DD} \gg 1$, for most of the parameter space we consider), long-range new force between dark matter particles necessarily significantly modifies the structure of the dark matter halo itself, as we discuss in Section \ref{sub:UFDChanges}.

The exception to this is for dark matter with zero net charge (e.g. equal numbers of particles with opposite charges), in which case dynamical friction remains present (since it requires a net charge for stars, but not for dark matter) but most such modifications to the halo itself do not (since there are no large net forces when the dark matter is net-neutral). However, in this case, one must account for the effects of Debye screening and plasma instabilities. We consider these in Section \ref{sub:PlasmaEffects}.

\subsection{Dynamical Friction with Net-Charged Dark Matter}\label{sub:UFDChanges}

In this subsection, we extend the gravity-only dynamical friction calculation of the previous subsection to models in which dark matter has a single charge, such that all dark matter particles attract one another or they all repel one another; we discuss example models leading to these two behaviors below. We leave the net-neutral case to the next subsection, and more general charge-asymmetric cases can be evaluated by combining the discussion in these two subsections. Note that the (Standard Model) stars we consider will never be net-neutral, since dynamical friction of stars would be negligible if they did not carry a large net charge under the new force.

In the single-charge case, our dynamical friction results must be modified (beyond the $\alpha_\SD^2$ scaling) by the modification of inferred UFD halo properties by the new Standard Model-dark matter and dark matter-dark matter forces. We begin by classifying the new forces we consider into four qualitative categories, each of which we will consider in turn below:
\begin{itemize}
    \item Forces that always attract dark matter particles both to other dark matter particles and to Standard Model particles (e.g.~a new scalar mediator coupled to both sectors, with the same sign of charge for both sectors).
    \item Forces that are always attractive among dark matter particles and always repulsive between dark matter and Standard Model particles (e.g.~a new scalar mediator coupled to both sectors, with opposite charge signs between the two sectors).
    \item Forces that are always repulsive among dark matter particles and always attractive between dark matter and Standard Model particles (e.g.~a massive vector arising from $U(1)_{B-L}$, with all dark matter particles carrying negative $B-L$ charge while stars' neutrons make them positive).
    \item Forces that always repel dark matter particles both from other dark matter particles and from Standard Model particles (e.g. a massive vector arising from $U(1)_{B-L}$, with both sectors carrying positive $B-L$ charge).
\end{itemize}

Introducing a force between dark matter particles would modify the condition for hydrostatic equilibrium in the halo, Eq.~\eqref{eq:HydrostaticEquilibriumD}, requiring a change in the total mass, scale radius (keeping the profile shape fixed), and/or temperature distribution of the halo in order to preserve equilibrium. Similarly, introducing a force between dark matter particles and Standard Model particles affects the condition for the stars' hydrostatic equilibrium, Eq.~\eqref{eq:HydrostaticEquilibriumS}. The latter, however, can be resolved only by adjusting the central dark matter density of the profile, since the stars' temperature distribution is (roughly) fixed by their observed velocity dispersion. For any chosen profile shape (i.e. $\gamma=0$, $\gamma=1$, or NFW), there are two ways to change this central density: a change in the total mass of the halo, or a change in its scale radius (or a combination of the two).

For the purposes of estimates in this work, we will assume that the scale radius of Segue I's halo is fixed to its gravity-only value, and will instead vary the halo's total mass in order to preserve the known stellar velocity dispersion, while varying the dark matter temperature distribution in order to preserve its hydrostatic equilibrium. This is, of course, inexact: re-running structure formation simulations in the presence of these significant long-ranged interactions would presumably result in different scale radii for UFDs as well. However, such a change cannot be all that large since Segue I's dark matter scale radius is bounded from below by the stellar scale radius ($\sim30$ pc) and from above by its distance from the Milky Way center ($\sim30$ kpc \cite{Segue1Location}). Propagating the effects of increasing the assumed $R_\chi$ by a factor of $N$ through our calculation gives a change of only $\sim N$ in the heating rate for $\gamma=0$ and $\gamma=1$ profiles, or $\sim N^{3/2}$ for NFW profiles. This is not particularly significant for our rough estimates, considering the quadratic (up to corrections discussed below) dependence of the heating rate on $\alpha_\SD$, unless one is willing to assume that Segue I's size is a large fraction of its distance from the Milky Way center. Assuming that the scale radius is fixed and varying only the mass and temperature distribution of $\chi$ when introducing new forces is therefore sufficient for our order-of-magnitude estimates.

Before we evaluate these changes in the mass and temperature distribution of Segue I's dark matter, it is helpful to define effective force strengths $\alpha_\SD^\eff$ and $\alpha_\DD^\eff$ by
\begin{subequations}\begin{align}
    \alpha_\SD^\eff &\sim \frac{\alpha_\SD}{1+(R_\chi/\lambda)^2} \qquad {\rm for~}\gamma=0 \\
    \alpha_\DD^\eff &\sim \frac{\alpha_\DD}{1+(R_\chi/\lambda)^2} \qquad {\rm for~}\gamma=0
\end{align}\end{subequations}
and
\begin{subequations}\begin{align}
    \alpha_\SD^\eff &\sim \frac{\alpha_\SD}{1+(R_*/\lambda)^2} \qquad {\rm for~NFW~or~}\gamma=1 \\
    \alpha_\DD^\eff &\sim \frac{\alpha_\DD}{1+(R_*/\lambda)^2} \qquad {\rm for~NFW~or~}\gamma=1.
\end{align}\end{subequations}
These give the actual strength of the new force seen by a particle in the core of the UFD, accounting for the near-uniform density of dark matter on scales $\lambda \lesssim R_*$, which weakens the resulting new force in this regime. For a derivation of this suppression, see Ref. \cite{LRDMSI}, noting that we are interested in the forces at $R_* \ll R_\chi$.

We can now evaluate the changes to Segue I's halo properties, and thus the changes to our dynamical friction calculation, in the presence of new forces. We begin with universally attractive forces. In this case, the density of dark matter in the halo must be reduced by a factor of $1+\alpha_\SD^\eff$, in order to maintain the observed stellar velocity dispersion. This in-turn reduces the dark matter velocity dispersion by a factor of $\sqrt{1+\alpha_\SD^\eff}$ (see Eq. \eqref{eq:SigmaDRS}), but there is a competing increase in $\chi$'s velocity caused by the new dark matter-dark matter force; this enhancement is given by $\sqrt{1+\alpha_\DD^\eff}$.

At this point we see an additional complication: we wish to constrain $\alpha_\SD$, but our result depends on $\alpha_\DD$ as well. Since the heating rate Eqn.~\eqref{eq:TotalHeatingRate} is, for $T_\chi \ll T_*$, proportional to $M_\chi\sigma_\chi^{-3/2}$ until $\sigma_\chi = \sigma_*$ (after which it is independent of $\sigma_\chi$), the weakest constraint on $\alpha_\SD$ will arise when $\alpha_\DD$ is equal to the maximum value allowed by other observations, $\alpha_\DD^\MAX$. The time needed for the stellar scale radius to reach its present value, starting from the dark matter scale radius, is thus
\begin{align}
    \tau_{\rm eq} &\sim  \tau_{\rm eq}^G \alpha_\SD^{-2} \frac{\ln \Lambda_G}{\ln \Lambda_Y} \times \begin{cases} \frac{\left(1+\alpha_\DD^\MAX\right)^{3/2}}{\sqrt{1+\alpha_\SD^\eff}} & \sigma_\chi \gtrsim \sigma_* \\ \pfrac{\sigma_{*,0}}{\sigma_{\chi,0}}^{3/2} \left(1+\alpha_\SD^\eff\right) & \sigma_\chi \lesssim \sigma_* \end{cases} \label{eq:AttAttSolution}
\end{align}
where $\Lambda_G$ is the Coulomb logarithm for gravity (Eqn.~\eqref{eq:CoulombLog}) while $\Lambda_Y$ is the analogous logarithm for a Yukawa force (see Eq.~\eqref{eq:YukawaLog} in Appendix \ref{app:FiniteRangeHeating}) and $\sigma_{*,0}$ ($\sigma_{\chi,0}$) are the velocity dispersions of the stars (of $\chi$) calculated assuming no new forces; $\Lambda_Y$ implicitly depends on $\alpha_\SD$ and $\lambda$. Note the presence of both $\alpha_\SD$ and $\alpha_\SD^\eff$ in Eq. \eqref{eq:AttAttSolution}: while the reduction of the dark matter density is suppressed for $R_\chi \gg \lambda$, dynamical friction is not. 

\begin{figure}
    \centering
    \includegraphics[width=0.7\linewidth]{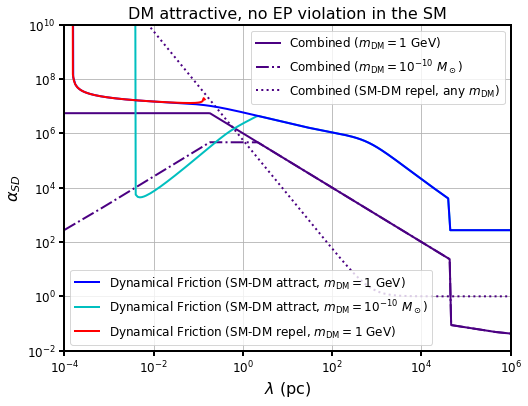}
    \caption{The upper bounds on the strength $\alpha_\SD$ of a new force as a function of its range $\lambda$, see Eq.~\eqref{eq:YukawaSD} (with $q_S=q_D=1$). This figure covers the case of an attractive force between dark matter (DM) and either an attractive or repulsive force between dark matter and Standard Model (SM) particles, in the limit of zero equivalence principle (EP) violation arising from the SM-SM force. Shown are the bounds from combining constraints on DM-DM interactions with constraints on SM-SM interactions we present in Section \ref{sec:CombinedConstraints}, as well as the bounds from dynamical friction in ultrafaint dwarf galaxies (UFDs) that we derive in Section \ref{sec:DFinUFDs}. The non-dynamical friction bounds are shown for two example DM masses $m_\DM$, along with a DM mass-independent bound (arising from the need for stars to stay bound to UFDs) that applies only if DM and SM particles repel one another. The dynamical friction bounds are likewise shown for both DM masses and for both the attractive and repulsive SM-DM cases, although the distinction between the latter two is small (except that the repulsive case is nonsensical beyond the UFD stellar binding line). Results are shown only for a $\gamma=0$ profile; no constraint can be obtained for the $\gamma=1$ or NFW profiles when the new force attracts dark matter to itself.}
    \label{fig:AttractiveNEPV}
\end{figure}

\begin{figure}
    \centering
    \includegraphics[width=0.7\linewidth]{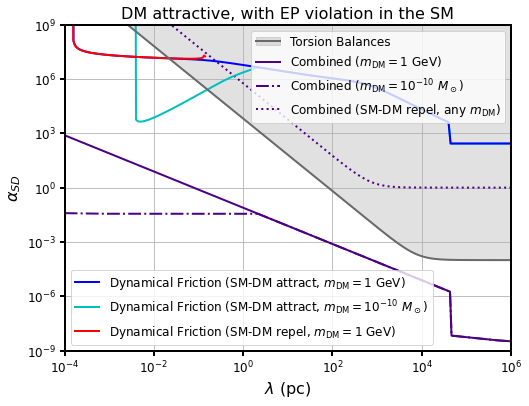}
    \caption{Similar to Fig.~\ref{fig:AttractiveNEPV} but here with EP violation in the SM-SM force.  This figure shows the upper bounds on new forces that attract dark matter (DM) particles to one another and either attract or repel dark matter to/from Standard Model (SM) particles, in terms of their range $\lambda$ and their strength relative to gravity $\alpha_\SD$---see Eq. \eqref{eq:YukawaSD} (with $q_S=q_D=1$)---in the limit of maximal equivalence principle (EP) violation arising from the SM-SM force. Other than the change in bounds on the SM-SM force in the EP-violating case, and the addition of the torsion balance-based test of Milky Way center-directed equivalence principle violation in Ref. \cite{AlphaSDDirectTorsion}, this plot is identical to Fig.~\ref{fig:AttractiveNEPV}; see that figure's caption for details.}
    \label{fig:AttractiveWEPV}
\end{figure}

Requiring that $\tau_{\rm eq} \gtrsim 10$ Gyr thus gives us an implicit solution for our constraint. This is plotted, for each of the profile shapes we consider, in Figs. \ref{fig:AttractiveNEPV} and \ref{fig:AttractiveWEPV} (including two example dark matter masses in each, since the mass of $\chi$ affects the value of $\alpha_\DD^\MAX$). The two figures differ in their assumptions about the existing constraints on Standard Model-Standard Model forces; see Fig. \ref{fig:IndividualConstraints} and the discussion in Section \ref{sec:CombinedConstraints}. We discuss some features of these figures below, after handling the remaining cases.

For forces that are attractive among $\chi$ particles but repulsive between $\chi$ and stars, $\alpha_\SD^\eff$ now increases the inferred halo mass rather than decreasing it; this changes a sign in Eq. \eqref{eq:AttAttSolution} and eliminates the need to consider $\sigma_\chi \lesssim \sigma_*$:
\begin{align}
    \tau_{\rm eq} &\sim  \tau_{\rm eq}^G \alpha_\SD^{-2} \frac{\ln \Lambda_G}{\ln \Lambda_Y} \times \frac{\left(1+\alpha_\DD^\MAX\right)^{3/2}}{\sqrt{1-\alpha_\SD^\eff}}. \label{eq:AttRepSolution}
\end{align}
Note that there is also a straightforward bound on $\alpha_\SD$ in this case arising from the requirement that UFD's stars are bound: $\alpha_\SD^\eff < 1$. Both of these bounds are plotted in Figs. \ref{fig:AttractiveNEPV} and \ref{fig:AttractiveWEPV}.

\begin{figure}
    \centering
    \includegraphics[width=0.7\linewidth]{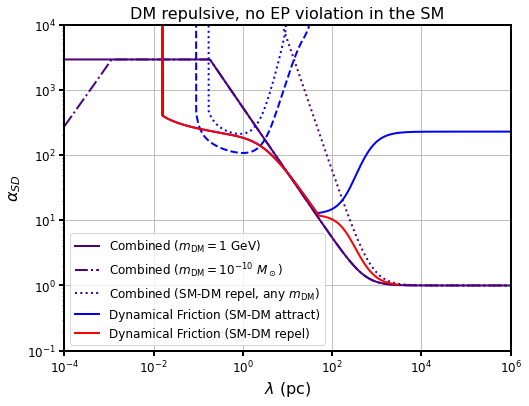}
    \caption{The upper bounds on new forces that repel dark matter (DM) particles from one another and either attract or repel dark matter to/from Standard Model (SM) particles, in terms of their range $\lambda$ and their strength relative to gravity $\alpha_\SD$---see Eq.~\eqref{eq:YukawaSD} (with $q_S=q_D=1$)---in the limit of zero equivalence principle (EP) violation arising from the SM-SM force. Shown are the bounds from combining constraints on DM-DM interactions with constraints on SM-SM interactions we present in Section \ref{sec:CombinedConstraints}, as well as the bounds from dynamical friction in ultrafaint dwarf galaxies (UFDs) that we derive in Section \ref{sec:DFinUFDs}. The non-dynamical friction bounds are shown for two example DM masses $m_\DM$, along with a DM mass-independent bound (arising from the need for stars to stay bound to UFDs) that applies only if DM and SM particles repel one another. The dynamical friction bounds are, in this case, independent of DM mass; see the discussion above Eq. \eqref{eq:RepAttSolution}. We show the dynamical friction bounds for both the attractive and repulsive SM-DM cases, although the distinction between the latter two is small (except that the repulsive case is nonsensical beyond the UFD stellar binding line). To illustrate our sensitivity to the choice of assumed UFD halo profile, the attractive cases are shown for $\gamma=0$ (solid), $\gamma=1$ (dotted), and NFW (dashed) profiles.}
    \label{fig:RepulsiveNEPV}
\end{figure}

\begin{figure}
    \centering
    \includegraphics[width=0.7\linewidth]{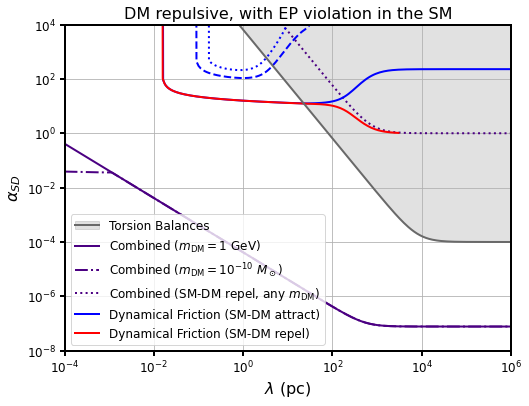}
    \caption{The upper bounds on new forces that repel dark matter (DM) particles from one another and either attract or repel dark matter to/from Standard Model (SM) particles, in terms of their range $\lambda$ and their strength relative to gravity $\alpha_\SD$---see Eq. \eqref{eq:YukawaSD} (with $q_S=q_D=1$)---in the limit of maximal equivalence principle (EP) violation arising from the SM-SM force. Other than the change in bounds on the SM-SM force in the EP-violating case (which then affect our bounds for reasons discussed around Eqs. \eqref{eq:RepAttSolution} and \eqref{eq:RepRepSolution}), and the addition of the torsion balance-based test of Milky Way center-directed equivalence principle violation in Ref. \cite{AlphaSDDirectTorsion}, this plot is identical to Fig. \ref{fig:RepulsiveNEPV}; see that figure's caption for details.}
    \label{fig:RepulsiveWEPV}
\end{figure}

Situations where dark matter particles repel one another can be handled similarly. In this case, the new dark matter self-interaction tends to slow down dark matter and thus enhance dynamical friction, so our bound must instead use the smallest possible $\alpha_\DD$. For a given value of $\alpha_\SD$, this is given by $\alpha_\SD^2/\alpha_\SMSM^\MAX$, where $\alpha_\SMSM^\MAX$ is the largest value of $\alpha_\SMSM$ allowed at the given range. If dark matter and stars attract, we thus have, 
\begin{align}
    \tau_{\rm eq} &\sim  \tau_{\rm eq}^G \alpha_\SD^{-2} \frac{\ln \Lambda_G}{\ln \Lambda_Y} \times \begin{cases} \frac{\left(1-\frac{\alpha_\SD^\eff\alpha_\SD}{\alpha_\SMSM^\MAX}\right)^{3/2}}{\sqrt{1+\alpha_\SD^\eff}} & \sigma_\chi \gtrsim \sigma_* \\ \pfrac{\sigma_{*,0}}{\sigma_{\chi,0}}^{3/2} \left(1+\alpha_\SD^\eff\right) & \sigma_\chi \lesssim \sigma_* \end{cases} \label{eq:RepAttSolution}
\end{align}
which can again be numerically solved for a limit on $\alpha_\SD$. If dark matter and stars repel one another, we again change a sign:
\begin{align}
    \tau_{\rm eq} &\sim  \tau_{\rm eq}^G \alpha_\SD^{-2} \frac{\ln \Lambda_G}{\ln \Lambda_Y} \times \begin{cases} \frac{\left(1-\frac{\alpha_\SD^\eff\alpha_\SD}{\alpha_\SMSM^\MAX}\right)^{3/2}}{\sqrt{1-\alpha_\SD^\eff}} & \sigma_\chi \gtrsim \sigma_* \\ \pfrac{\sigma_{*,0}}{\sigma_{\chi,0}}^{3/2} \left(1-\alpha_\SD^\eff\right) & \sigma_\chi \lesssim \sigma_* \end{cases} \label{eq:RepRepSolution}
\end{align}
where we have kept both cases because $\sigma_\chi \lesssim \sigma_*$ may be possible, depending on $\alpha_\SMSM^\MAX$. The bounds for these two cases are plotted in Figs. \ref{fig:RepulsiveNEPV} and \ref{fig:RepulsiveWEPV}, where the difference is again the assumptions made about EP violation in the Standard Model-Standard Model force.

Several features in Figs. \ref{fig:AttractiveNEPV}--\ref{fig:RepulsiveWEPV} are readily understood. First, note the varying minimal ranges at which any constraint can be set. If not for the modification of inferred halo properties in the presence of new forces, all of the exclusion lines would end at a range of order mpc (depending on the profile shape) because the momentum-transfer cross-section is essentially saturated once every $\chi$ particle within $\lambda$ of a star is scattered by order-one; raising the coupling further has only a logarithmic effect on the heating rate\footnote{For simplicity, we conservatively omit this logarithmic effect and omit scattering at ranges greater than $\lambda$ entirely; see Eq. \eqref{eq:YukawaCutoff} and the discussion that follows it}. This minimum range would thus correspond to a total mass of dark matter within $\lambda$ equal to the star's mass, i.e
\begin{align}
    \lambda_{\rm min}^\text{(na\"{i}ve)} &\sim \sqrt{\frac{M_\odot}{\pi v_{\rm th}\times10~\text{Gyr}\times\rho_\chi}}.
\end{align}

Accounting for the modification of inferred halo properties in the presence of new forces complicates this story. In particular, $\rho_\chi$ differs from its gravity-only value whenever $\alpha_\SD^\eff \gtrsim 1$, and $v_{\rm th}$ is changed whenever $\alpha_\DD^\eff \gtrsim 1$ (for both attractive and repulsive forces). This changes the minimal $\lambda$ values where any $\alpha_\SD$ is excluded, but the general picture is unchanged.

Second, it is noticeable that $\gamma=0$ profiles generally give substantially tighter constraints than the $\gamma=1$ and NFW profiles. (In fact, when dark matter attracts itself, $\gamma=1$ and NFW profiles lead to no constraints at all.) This results simply from the much more rapid decrease in dark matter density in the two cuspy profiles than in the cored $\gamma=0$ profile: since the known stellar velocity distribution forces all three profiles to have similar $\rho_\chi$ around $R_*$, the cuspy profiles have much smaller dark matter densities at larger scale radii\footnote{Note that this effect relies significantly on the dark matter density scaling as $\rho_\chi \propto r^{-1}$ for radii as small as $R_*$. A profile whose density increases as one decreases $r$ below $R_\chi$, but whose density levels off at a radius larger than $R_*$, would give results somewhere between our cuspy and cored examples.}. This effect can be seen for the gravity-only case in Fig. \ref{fig:TimeEvolutionDF}. As we discuss in Section \ref{sub:AlphaAlphaConstraints}, Figs. \ref{fig:AlphaAlphaAtt}--\ref{fig:AlphaAlphaNN} also show constraints for the $\gamma=1$ and NFW profiles requiring stars to start within $10R_*$, instead of within $R_\chi$ (for $\gamma=0$ we have $R_\chi < 10R_*$, so this isn't sensible); this substantially strengthens these constraints, not only because of the reduced deorbiting required, but also because it avoids the very low dark matter densities at large radii.

Most of the other sharp features in these results come from transitions between the various constraints on $\alpha_\DD$ and $\alpha_\SMSM$, due to their downstream effects on the most conservative halo assumptions. The exceptions correspond to situations when one of the halo correction factors above (e.g. when the factors of $1\pm\alpha_\SD^\eff$ in Eqs.~\eqref{eq:AttAttSolution}--\eqref{eq:RepRepSolution}) change from being essentially equal to $1$ to varying significantly; see, for example, the sudden separation of the SM-DM attracting and repelling lines in Fig. \ref{fig:RepulsiveNEPV} above approximately 30 pc, compared to their essentially identical values at lower ranges.

Finally, while most of our constraints apply to new forces with $\alpha_\SD \gg 1$ --- unsurprisingly, since gravity alone leads to negligible dynamical friction --- a few exceptions can be seen at large ranges in Figs. \ref{fig:RepulsiveNEPV} and \ref{fig:RepulsiveWEPV}. These are nominally strong constraints, but they correspond to fairly contrived scenarios, in which the mutual repulsion of dark matter almost exactly cancels the effect of gravity, leading to extremely low dark matter velocities, and consequently dramatically enhanced dynamical friction. Thus, while these ``strong'' constraints are correct, they are unlikely to apply to many models of interest.

During the final preparation of this work, Ref. \cite{OtherDynamicalFrictionBound} was released and presented a dynamical friction constraint on $\alpha_\SD$ similar to ours. However, there are several significant differences between that work and ours: First, we consider more general classes of new forces, including not only universally attractive forces but also repulsive forces and forces under which dark matter is net-neutral. Second, we assume throughout our work that a new Standard Model-dark matter interaction necessarily implies new Standard Model-Standard Model and dark matter-dark matter forces. For much of the relevant parameter space, the resulting dark matter-dark matter interaction meaningfully changes the inferred properties of UFDs, significantly modifying the dynamical friction heating rate. This effect is absent in Ref. \cite{OtherDynamicalFrictionBound}, which considers a force between the Standard Model and dark matter in isolation. Third, we more carefully handle the non-perturbative interaction regime, in which all dark matter particles within $\lambda$ of a star transfer an order-one fraction of their momentum, at which point further increases in the interaction strength no longer increase the dynamical friction rate (see Appendix \ref{app:FiniteRangeHeating}). This saturation is not present in Ref. \cite{OtherDynamicalFrictionBound}, resulting in that work's bounds extending to ranges somewhat too small for their assumptions to hold. However in those cases where there is overlap and their calculations are valid, our results are in good agreement with theirs.

\subsection{Dynamical Friction with Net-Neutral Dark Matter --- Plasma Effects}\label{sub:PlasmaEffects}

When dark matter is net-neutral---for example, in the case of millicharged dark matter with equal amounts of positive and negative charge---the modifications of inferred halo properties discussed in the previous section no longer apply, since there are no net forces on or by the dark matter collectively. However, dynamical friction can still be significant, since it is independent of the sign of the charge (note the quadratic dependence on $\alpha_\SD^2$ in Eq. \eqref{eq:TotalHeatingRate}). 

Dynamical friction was originally derived and is used extensively in the context of attractive forces such as gravity, and, to our knowledge, has not been discussed for a charged object moving in a net-neutral plasma. However, an alternative and equivalent picture of stellar infall can be derived purely from two-to-two scattering and the resultant heat transfer, as originally shown in Ref.~\cite{DFofStarsinUFDs}. This picture makes it clear that this effect is additive in the square of the individual charge and hence applies to net neutral plasmas as well.

Two complications relative to dynamical friction from gravity alone are present in this case: First, the presence of the dark matter plasma leads to Debye screening of the force, effectively reducing the force's range. Second, as we noted in Section \ref{sec:CombinedConstraints}, dark matter plasmas may be constrained in the near future by the emergence of two-stream instabilities in various astrophysical systems \cite{DMPlasmaInsabilities}. As we describe below, dynamical friction cannot set new constraints in this parameter space on its own, but it may be able to in combination with these potential plasma instability constraints.

The Debye screening length for forces in this case is given by the usual result,
\begin{align}
    \lambda_D^2 = \frac{T_\chi}{4\pi (\alpha_\DD G m_\chi^2) n_\chi} \sim \frac{\sigma_\chi^2}{4\pi \alpha_\DD G \rho_\chi},
\end{align}
with $n_\chi$ the number density of $\chi$ and $T_\chi \sim m_\chi\sigma_\chi^2$ as before. The (inverse) Debye length can thus be treated as an $\alpha_\DD$-dependent increase to the mediator mass, since both lead to exponential suppression of the force at long ranges.

This dependence on $\alpha_\DD$, as opposed to $\alpha_\SD$, forces us to choose the most conservative---i.e.~the largest---allowed $\alpha_\DD$ when determining whether a particular $\alpha_\SD$ is excluded, just as we did in some cases in the last section. As we noted previously, however, essentially no constraints on $\alpha_\SD$ can be set using only existing constraints on $\alpha_\DD$, as Debye screening renders new forces too short-ranged for dynamical friction to matter. (See also Fig. \ref{fig:AlphaAlphaNN} and the discussion in Section \ref{sub:AlphaAlphaConstraints}.) We therefore instead present the constraints that would be present if exponentially-growing plasma instabilities in dark matter, as described in Ref. \cite{DMPlasmaInsabilities}, are excluded as well; this leads to non-trivial results, and may be the case in the foreseeable future.

\begin{figure}
    \centering
    \includegraphics[width=0.7\linewidth]{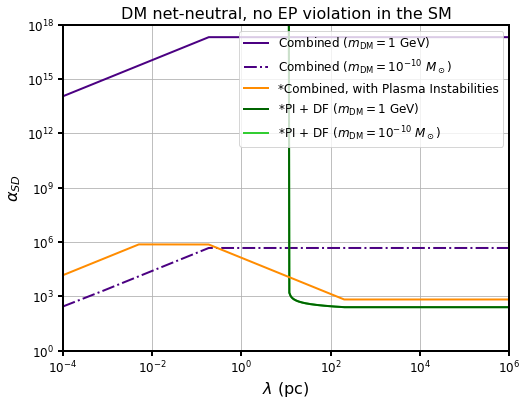}
    \caption{The bounds on new forces when dark matter is net-neutral but stars are not, in terms of their range $\lambda$ and their strength relative to gravity $\alpha_\SD$---see Eq. \eqref{eq:YukawaSD} (with $q_S=q_D=1$)---in the limit of zero equivalence principle (EP) violation arising from the SM-SM force. Shown are the existing bounds from combining constraints on DM-DM interactions with constraints on SM-SM interactions we present in Section \ref{sec:CombinedConstraints}, as well as the analogous bounds that would be obtained if exponentially growing plasma instabilities in astrophysical systems were excluded in the future; see Ref. \cite{DMPlasmaInsabilities}. Also shown (``PI+DF'') are the bounds from dynamical friction in ultrafaint dwarf galaxies (UFDs) that we derive in Section \ref{sec:DFinUFDs}, again assuming the plasma instability constraints; no meaningful constraints can be obtained without this assumption, due to the effects of Debye screening. The dynamical friction line is for $\gamma=0$ (solid), as no constraint on $\alpha_\SD$ can be set for the $\gamma=1$ and NFW profiles (but see Section \ref{sub:AlphaAlphaConstraints}). Two example DM masses $m_\DM$ are included (though they are not relevant for the plasma instability line alone). For dynamical friction, $m_\DM$ affects the existing constraints on $\alpha_\DD$ and thus the maximum strength of Debye screening, although this is only visible for NFW profiles; see Sec. \ref{sub:PlasmaEffects}.}
    \label{fig:NeutralNEPV}
\end{figure}

\begin{figure}
    \centering
    \includegraphics[width=0.7\linewidth]{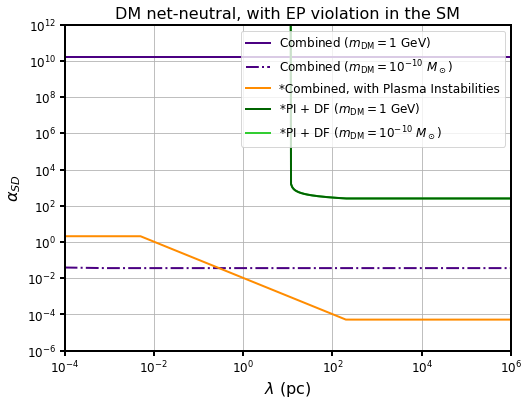}
    \caption{The bounds on new forces when dark matter is net-neutral but stars are not, in terms of their range $\lambda$ and their strength relative to gravity $\alpha_\SD$---see Eq. \eqref{eq:YukawaSD} (with $q_S=q_D=1$)---in the limit of zero equivalence principle (EP) violation arising from the SM-SM force. Other than the change in bounds on the SM-SM force in the EP-violating case, this plot is identical to Fig. \ref{fig:NeutralNEPV}; see that figure's caption for details.}
    \label{fig:NeutralWEPV}
\end{figure}

From there, constraints can be calculated in the same way as before, evaluating the minimum value of $\alpha_\SD$ for which $|\tau_{\rm eq}| \lesssim 10$ Gyr at a given $\lambda$. The results are plotted in Figs. \ref{fig:NeutralNEPV} and \ref{fig:NeutralWEPV}, once again split between the limits of zero and maximal EP violation on the Standard Model side. As those figures illustrate, dynamical friction in UFDs cannot exclude any new values of $\alpha_\SD$ for $\gamma=1$ and NFW profiles, even if plasma instabilities are excluded in the future. It may, however, still be able to exclude portions of the larger new force parameter space; this is the discussed in the next section.

\subsection{\texorpdfstring{Constraints on $(\alpha_\SMSM, \alpha_\DD)$}{Constraints on (alpha\_SS, alpha\_DD)}}\label{sub:AlphaAlphaConstraints}

\begin{figure}
    \centering
    \includegraphics[width=0.7\linewidth]{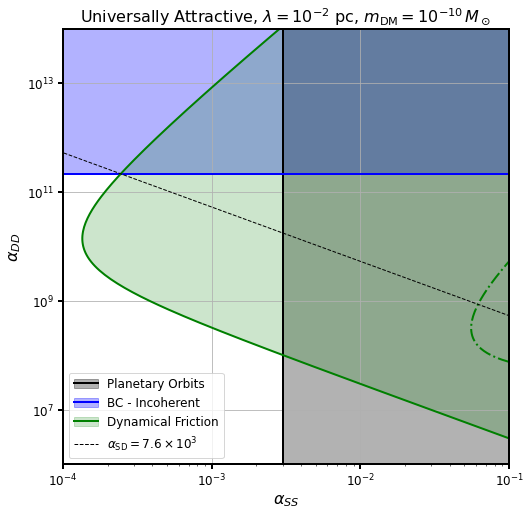}
    \caption{Constraints on the $(\alpha_\SMSM, \alpha_\DD)$  plane for a universally attractive (i.e.~always attractive for both Standard Model and dark matter particles) force, at a fixed range of $\lambda = 10^{-2}$ pc and assuming a dark matter mass of $m_\DM = 10^{-10}\,M_\odot$. Shown are the constraint on $\alpha_\SMSM$ from observations of planetary orbits \cite{AlphaSSnEPVPlanets, AlphaSSBook, AlphaSSReview} (the constraint on equivalence principle-violation Standard Model forces is off the left edge of the plot; see Fig. \ref{fig:IndividualConstraints}) and the constraint on $\alpha_\DD$ from incoherent scattering in the bullet cluster, as well as the excluded region from our dynamical friction calculation for the $\gamma=0$ profile (green solid line). The green dot-dashed line shows the constraint that would be obtained for both $\gamma=1$ and NFW profiles (up to a negligible difference) if we required the initial stellar scale radius to be $\leq 10R_*$, rather than $\leq R_\chi$; no constraints would appear on this plot for $\gamma=1$ or NFW profiles using our usual assumptions. The black dashed line illustrates the $\alpha_\SD \lesssim 4.4\times10^2$ constraint arising from a $\gamma=0$ profile. The direct constraint on $\alpha_\SD$ from torsion balances found in Ref. \cite{AlphaSDDirectTorsion} is above and to the right of this plot's boundaries.}
    \label{fig:AlphaAlphaAtt}
\end{figure}

\begin{figure}
    \centering
    \includegraphics[width=0.7\linewidth]{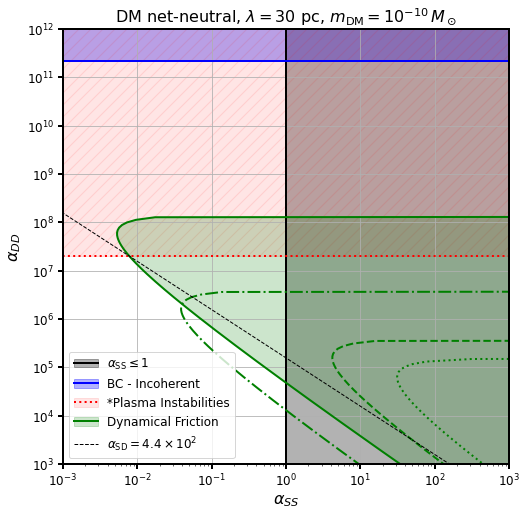}
    \caption{Constraints on the $(\alpha_\SMSM, \alpha_\DD)$ plane when dark matter is net-neutral, at a fixed range of $\lambda = 10^3$ pc and assuming a dark matter mass of $m_\DM = 10^{-10}\,M_\odot$. Shown are the constraint on $\alpha_\SMSM$ from the known strength of gravity (the constraint on equivalence principle-violating Standard Model forces is off the left edge of the plot; see Fig. \ref{fig:IndividualConstraints}) and the constraint on $\alpha_\DD$ from incoherent scattering in the bullet cluster, as well as the excluded regions from our dynamical friction calculations assuming the usual three profile shapes: $\gamma=0$ (solid), $\gamma=1$ (dotted), and NFW (dashed). The green dot-dashed line shows the constraint that would be obtained for both $\gamma=1$ and NFW profiles (up to a negligible difference) if we required the initial stellar scale radius to be $\leq 10R_*$, rather than $\leq R_\chi$. Also shown is the potential future constraint from plasma instabilities in astrophysical systems. Without assuming that the plasma instability region is excluded, no constraint on $\alpha_\SD$ can be set by dynamical friction except for NFW profiles (although part of the $(\alpha_\SMSM, \alpha_\DD)$ plane is still excluded), due to the unexcluded parameter space at smaller $\alpha_\SMSM$ and larger $\alpha_\DD$ than the dynamical friction region. With that assumption, however, profile-dependent constraints can be set: the black dashed line, for example, illustrates the $\alpha_\SD \lesssim 7.6\times10^3$ constraint arising from a $\gamma=0$ profile.}
    \label{fig:AlphaAlphaNN}
\end{figure}

Over the last several sections, we have focused on constraints on $\alpha_\SD$ as a function of $\lambda$, but this is only a two-dimensional projection of the $(\lambda, \alpha_\SMSM, \alpha_\DD)$ parameter space. This understates the exclusion potential of dynamical friction, since we exclude values of $\alpha_\SD$ only if \textit{all} corresponding combinations of $(\alpha_\SMSM, \alpha_\DD)$ are excluded. In fact, there exists substantial parameter space where dynamical friction excludes new pairs of $(\alpha_\SMSM, \alpha_\DD)$ at a given $\lambda$, even if it cannot exclude the associated value of $\alpha_\SD = \sqrt{\alpha_\SMSM\alpha_\DD}$ because there exist non-excluded pairs $(\alpha_\SMSM, \alpha_\DD)$ which give the same $\alpha_\SD$. We illustrate examples of this in Figs. \ref{fig:AlphaAlphaAtt} and \ref{fig:AlphaAlphaNN}.

In Fig.~\ref{fig:AlphaAlphaAtt}, we show the $(\alpha_\SMSM, \alpha_\DD)$ parameter space for a fixed range of $\lambda = 10^{-2}$ pc when all the new forces are attractive. The dark matter mass is chosen to be $m_{\rm DM}=10^{-10} M_\odot$, setting the incoherent bound from bullet-cluster in blue, which is the strongest existing constraint on $\alpha_\DD$ for these parameters; the analogue of this plot for $m_{\rm DM} = 1$ GeV would be identical, except that the blue region would be located far above the top of the plot. In the limit of zero equivalence principle violation on the Standard Model side, the strongest constraint on $\alpha_\SMSM$ comes from planetary orbits, and is shown in black; constraints on forces with significant equivalence principle violation are off of the left edge of the plot. The green curves illustrate the constraints arising from dynamical friction caused by the new Standard Model-dark matter force for the $\gamma=0$ profile (with our standard assumptions), as well as for an NFW profile with one assumption modified: the initial stellar scale radius is required to be $\leq 10~R_*$, rather than $\leq R_\chi$. (The $\gamma=1$ line with this assumption is essentially identical to this NFW line.) The NFW and $\gamma=1$ curves would be off the plot entirely using our standard, more conservative assumption. The thin, black dotted line, conversely, shows the value of $\alpha_\SD$ that we can exclude in the $\gamma=0$ case. Note that this significantly understates the exclusion potential of dynamical friction: there is a large, newly-excluded (in the limit of zero equivalence principle violation) region below this line, but this fact is obscured when considering limits on $\alpha_\SD$ alone.

In Fig. \ref{fig:AlphaAlphaNN}, we show the $(\alpha_\SMSM, \alpha_\DD)$ parameter space for a fixed range of $\lambda = 30$ pc when dark matter is net-neutral. Existing constraints from incoherent scattering of dark matter in the bullet cluster assuming $m_{\rm DM}=10^{-10}M_\odot$ are shown in blue. Constraints from the known strength of gravity are shown in black, while the red hatched region illustrates where plasma instabilities in astrophysical systems will grow exponentially. The dynamical friction constraints arising from $\alpha_\SD = \sqrt{\alpha_\SMSM\alpha_\DD}$ are shown in green, including our usual three profile lines and one addition: a dot-dashed line illustrating the (essentially identical) results for $\gamma=1$ and NFW profiles when requiring the initial stellar scale radius to be $\leq 10R_*$, rather than $\leq R_\chi$. We see that dynamical friction constrains significant new parameter space in this plane for $\gamma=0$ profiles, or with this slightly less conservative assumption. However, when converting this to limits in the $(\lambda,\alpha_\SD)$ plane, we cannot yet set any constraint as there is parameter space at larger $\alpha_\DD$ and smaller $\alpha_\SMSM$, giving the same $\alpha_\SD$, that is not presently excluded. The hypothetical constraint from plasma instabilities would exclude this large $\alpha_\DD$ region, allowing dynamical friction with $\gamma=0$ to constrain $\alpha_\SD$. This is why we only show constraints assuming plasma instabilities are excluded in Section \ref{sub:PlasmaEffects}.

\section{Conclusion}\label{sec:Conclusion}

In this work, we have presented two classes of constraints on long-range forces between Standard Model particles and dark matter. First, constraints on $\alpha_\SD$ can be obtained by combining constraints on $\alpha_\SMSM$ and $\alpha_\DD$ separately (Eq.~\eqref{eq:GeometricMean}), since any new SM-DM force also gives a new SM-SM force and an new DM-DM force. To that end, we have reviewed the strongest existing constraints on these two couplings in order to present the strongest available combined constraints on $\alpha_\SD$. These combined constraints are, in most models, stronger than any constraints that can be placed directly on $\alpha_\SD$.

Second, we derived new constraints on $\alpha_\SD$ from the effects of dynamical friction in ultrafaint dwarf galaxies. We showed that the introduction of new long-ranged forces between dark matter and the Standard Model could accelerate the transfer of energy from stars to their dark matter halos, potentially resulting in those stars necessarily deorbiting to smaller-than-observed radii over these UFD's lifetimes. We presented the resulting constraints subject to a variety of assumptions about the new force and about the UFD's halo profile. 

If the DM is net-charged there are essentially eight cases: the DM-DM force can be either attractive or repulsive, the DM-SM force can be either attractive or repulsive, and the SM-SM force can be either equivalence principle violating or not\footnote{The amount of equivalence principle violation is really a continuum of course, but we present the two limits.}. Our main results for these eight cases are presented in Figures \ref{fig:AttractiveNEPV}--\ref{fig:RepulsiveWEPV} and \ref{fig:AlphaAlphaAtt}.  If instead the DM is a net neutral plasma, our main results are presented in Figures \ref{fig:NeutralNEPV}--\ref{fig:NeutralWEPV} and \ref{fig:AlphaAlphaNN}.

The constraints we derive from dynamical friction cover new parameter space in the $(\lambda,\alpha_\SMSM, \alpha_\DD)$ space over and above existing limits on these individual couplings, for some assumptions about the force's and UFD halo's properties. Translating these constraints to constraints on $\alpha_\SD$ is non-trivial, however. For example, since dynamical friction weakens at large relative velocity, a large enough attractive $\alpha_\DD$ nullifies these constraints. As a result, the dynamical friction constraints are generally weaker than the constraints derived purely from the individual couplings in the $(\lambda, \alpha_\SD)$ plane, since one can dial up $\alpha_\DD$ and dial down $\alpha_\SMSM$ while keeping $\alpha_\SD$ fixed. Constraints from dynamical friction still set the best limits in this plane at some ranges for certain models, especially those where there is very little equivalence principle violation in the Standard Model. They are also the strongest \textit{direct} constraints on such forces in significant portions of the $(\lambda,\alpha_\SD)$ parameter space. 

It may be possible to meaningfully strengthen these constraints with improved astrophysical observations: observing objects with high central densities of dark matter but small halo radius (and thus low dark matter velocity dispersion) will generally strengthen our constraints. Similarly, these constraints would be strengthened by improved knowledge of existing objects: our constraints are limited significantly by poor knowledge of UFD dark matter halo profiles (e.g. we find much stronger constraints for cored profiles than cuspy ones) and by our highly conservative assumptions about star formation within them (namely that we assumed only that typical stars formed within a scale radius of the center). There is thus considerable potential for our analysis to give stronger constraints on $\alpha_\SD$ in the future.

\appendix

\section{Analytic Expressions for Halo Properties}\label{app:AnalyticExpressions}

In this appendix, we list the analytic expressions for various halo properties for $\gamma=1$ and NFW profiles, as we included only their $\gamma=0$ analogues in the main text for brevity.

We begin with the leading-order velocity dispersions at the stellar scale radius for $\gamma=1$,
\begin{align}
    \sigma_*(R_*) &= \sqrt{\frac{(8\sqrt{2}-10)GM_{\chi,\tot}R_*}{3R_\chi^2}} \\
    \sigma_\chi(R_*) &= \sqrt{\frac{GM_{\chi,\tot}R_*}{R_\chi^2}\ln\pfrac{R_\chi}{R_*}}
\end{align}
and for the NFW profile,
\begin{align}
    \sigma_*(R_*) &\approx \sqrt{\frac{0.12\,GM_{\chi,\tot}R_*}{R_\chi^2}};
\end{align}
the analogous expression for $\sigma_\chi(R_*)$ is extremely unwieldy. These results can be compared to Eqs. \eqref{eq:SigmaSRS} and \eqref{eq:SigmaDRS} in the main text.

We next present the contributions to the self-gravitational potential energy of halos with NFW and $\gamma=1$ profiles, compared to the $\gamma=0$ cases of Eqs. \eqref{eq:USD0} and \eqref{eq:UDS0}. For $\gamma=1$, these are
\begin{subequations}\begin{align}
    U_{*\chi} &= \frac{4GM_{*,\tot}M_{\chi,\tot}R_\chi^2}{\pi^{3/2}R_*^3}G^{3,3}_{3,3}\left(\frac{R_*^2}{R_\chi^2}\left|\begin{matrix}1 & \frac{3}{2} & 2 \\[2pt] 2 & \frac{5}{2} & \frac{5}{2}\end{matrix}\right.\right) \\
    U_{\chi*} &= \frac{4GM_{*,\tot}M_{\chi,\tot}R_\chi^2}{\pi^{3/2}R_*^3}G^{3,3}_{3,3}\left(\frac{R_*^2}{R_\chi^2}\left|\begin{matrix}1 & 1 & \frac{3}{2} \\[2pt] \frac{3}{2} & 2 & \frac{5}{2}\end{matrix}\right.\right)
\end{align}\end{subequations}
while the NFW result is
\begin{subequations}\begin{align}
    \begin{split}
    U_{*\chi} \approx\,& \frac{5\, GM_{*,\tot}M_{\chi,\tot}}{9R_*} \cdot \frac{(R_\chi^2/R_*)^2+\ln(1+R_\chi^2/R_*^2)}{(1+R_\chi^2/R_*^2)^{3/2}}
    \end{split}\\
    U_{\chi*} \approx\, & \frac{0.20\,GM_{*,\tot}M_{\chi,\tot}R_\chi^2}{R_*^3}G^{3,3}_{3,3}\left(\frac{R_*^2}{R_\chi^2}\left|\begin{matrix}1 & \frac{3}{2} & 2 \\[2pt] \frac{3}{2} & 2 & \frac{5}{2}\end{matrix}\right.\right)
\end{align}\end{subequations}
where we have kept only the leading-order terms (for $R_* \ll R_\chi$) in the latter two expressions. We omit $U_{**}$ here since it does not depend on the halo profile.

For the density overlap form factor, as defined in Eq. \eqref{eq:FormFactorDef}, we have, for $\gamma=1$,
\begin{align}
    \mathcal{F} &= \frac{2}{\pi^{5/2}}G^{3,3}_{3,3}\left(\frac{R_*^2}{R_\chi^2}\left|\begin{matrix}0 & \frac{1}{2} & 1 \\[2pt] 1 & \frac{3}{2} & \frac{5}{2}\end{matrix}\right.\right).
\end{align}
and, for the NFW profile,
\begin{align}
    \mathcal{F} &\approx 0.032\,G^{3,3}_{3,3}\left(\frac{R_*^2}{R_\chi^2}\left|\begin{matrix}\frac{1}{2} & 1 & 1 \\[2pt] 1 & \frac{3}{2} & \frac{5}{2}\end{matrix}\right.\right).
\end{align}

\section{Heat Transfer with Finite-Range Forces}\label{app:FiniteRangeHeating}

In this appendix, we generalize the Coulombic heating rate, Eq. \eqref{eq:TotalHeatingRate}, to finite-range forces. As in the main text, we take $m_\chi \ll m_*$ throughout this appendix, as this is the regime relevant to this work.

The center-of-mass frame differential cross-section for a Yukawa force with strength relative to gravity $\alpha$ and range $\lambda$ in the Born scattering regime is, to leading order,
\begin{align}
    \frac{d\sigma}{d\Omega} &= 4(\alpha Gm_*m_\chi)^2\frac{m_\chi^2}{(\lambda^{-2}+q^2)^2}
\end{align}
with $q$ the momentum transfer at the given angle. In the $\lambda \to \infty$ limit, this reduces to the familiar gravitational result,
\begin{align}
    \frac{d\sigma}{d\Omega} &\xrightarrow[\lambda\to\infty]{} \frac{\pi}{2}\frac{(\alpha Gm_*m_\chi)^2}{(\frac{1}{2}m_\chi v_{\rel}^2)^2(1-\cos\theta)^2},
\end{align}
with $v_\rel$ the relative velocity of the particles, which leads to the Coulombic heating rate in the main text. We are interested in the finite-$\lambda$ regime, however. In this case, we have
\begin{align}
    \frac{d\sigma_T}{d\Omega} &= 4(\alpha Gm_*m_\chi)^2\frac{m_\chi^2(1-\cos\theta)}{(\lambda^{-2}+2m_\chi^2v_{\rel}^2(1-\cos\theta))^2}
\end{align}
which can be integrated over the $\cos\theta = 1-\xi$ to $\theta=\pi$ range\footnote{A small-angle cutoff is needed to account for finite halo size, i.e. for there being a maximum physical impact parameter. One could also set a non-trivial large-angle cutoff if desired, but there is no need to here.} to give
\begin{align}\begin{split}
    \sigma_T = 2\pi & \pfrac{\alpha Gm_*}{v_{\rel}^2}^2 \left[\ln\pfrac{1+4m_\chi^2v_\rel^2\lambda^2}{1+2\xi m_\chi^2v_\rel^2\lambda^2} + \frac{1}{1+4m_\chi^2v_\rel^2\lambda^2} - \frac{1}{1+2\xi m_\chi^2v_\rel^2\lambda^2} \right]. \label{eq:YukawaSigmaTExact}
\end{split}\end{align}
This has well-behaved small- and large-$\lambda$ limits:
\begin{subequations}\begin{align}
    \sigma_T \xrightarrow[m_\chi v_\rel \lambda \gg 1]{}&~ 2\pi \pfrac{\alpha Gm_*}{v_{\rel}^2}^2 \ln\pfrac{4m_\chi^2v_\rel^2\lambda^2}{1+2\xi m_\chi^2v_\rel^2\lambda^2} \label{eq:YukawaSigmaTRelevant} \\
    \sigma_T \xrightarrow[m_\chi v_\rel \lambda \ll 1]{}&~ 4\pi \left(4-\xi^2\right) \left(\alpha Gm_*m_\chi^2\lambda^2\right)^2.
\end{align}\end{subequations}
Note that the latter limit corresponds to a point-like four-particle interaction rather different from the long-ranged forces we consider in this work, but we include it for completeness since it can be relevant for ultralight dark matter.

Our focus is on the former limit, which depends significantly on the small-angle cutoff $\xi$. To obtain an appropriate cutoff $\xi$, we start by conservatively setting a maximum impact parameter $b_{\rm max}$ of $b_{\rm max} \lesssim R_*$, since dark matter is interacting with many stars at once if one includes impact parameters larger than this. The scattering angle corresponding to this maximum impact parameter is then roughly given by the (exact) infinite-$\lambda$ result \cite{DFofStarsinUFDs}:
\begin{align}\begin{split}
    \xi = 1-\cos\theta_{\rm min} \approx&\, 1 - \frac{b_{\rm max}^2 v_\rel^4 - \alpha^2 G^2 m_*^2}{b_{\rm max}^2 v_\rel^4 + \alpha^2 G^2 m_*^2} \\
    =&\, \frac{2}{1+\pfrac{v_\rel^2 b_{\rm max}}{\alpha Gm_*}^2}. \label{eq:YukawaCutoff}
\end{split}\end{align} 
While this is not exact, it should be a reasonable approximation for $b_{\rm max} \lesssim \lambda$; excluding $b_{\rm max} \gg \lambda$ should never be significant since the force is exponentially suppressed at such ranges. (Put another way, setting the minimum angle to $\theta>0$, i.e. $\xi>0$, is only necessary in order to handle situations where $\lambda > R_*$ anyway, since otherwise large impact parameters $b>R_*$ are exponentially suppressed even without a cutoff.) We can therefore set $b_{\rm max} = \min(\lambda, R_*)$ without meaningfully affecting any results, and then safely use Eq. \eqref{eq:YukawaCutoff} with this more conservative $b_{\rm max}$.

We can now separate the $m_\chi v_\rel \lambda \gg 1$ result into three limiting regimes, depending on the value of $\xi$:
\begin{subequations}
\begin{numcases} {\sigma_T \sim }
    2\pi \pfrac{\alpha Gm_*}{v_{\rel}^2}^2 \ln\left(4m_\chi^2v_\rel^2\lambda^2\right) & for $\xi \ll \frac{1}{m_\chi^2 v_\rel^2 \lambda^2}$ \label{eq:YukawaSigmaT1} \\
    4\pi \pfrac{\alpha Gm_*}{v_{\rel}^2}^2 \ln\pfrac{v_\rel^2 b_{\rm max}}{\alpha Gm_*} & for $\frac{1}{m_\chi^2 v_\rel^2 \lambda^2} \ll \xi \ll 2$ \label{eq:YukawaSigmaT2} \\
    2\pi b_{\rm max}^2 & for $2-\xi \ll 1$ \label{eq:YukawaSigmaT3}
\end{numcases} \label{eqs:YukawaSigmaT}
\end{subequations}
The last of these is particularly interesting, since it corresponds to an $\alpha$-independent behavior where every dark matter particle within $b_{\rm max}$ is transferring an order-one fraction of its momentum already, such that further increases in the coupling do not affect the momentum-transfer cross section.

This behavior is closely related to another subtlety that we have not addressed so far: the limitations of the Born approximation. The Born approximation holds when \cite{BornApprox1, BornApprox2}
\begin{align}
    \alpha G m_* m_\chi^2 \lambda \ll 1.
\end{align}
Since we are working in the limit of $m_\chi v_\rel \lambda \gg 1$, this implies
\begin{align}
    \frac{\alpha Gm_*}{v_{\rel}^2} \ll \lambda
\end{align}
and thus $\sigma_T \lesssim \lambda^2$ in all three of the cases in Eq. \eqref{eqs:YukawaSigmaT} whenever the Born approximation holds.

Approximate analytic scattering cross-sections in the non-perturbative regime are available as well---see Refs.~\cite{BornApprox2, BeyondBorn1, BeyondBorn2, BeyondBorn3, BeyondBorn4}---but we do not need them for the estimates in this work: since growth of the momentum-transfer cross-section with coupling becomes extremely slow once the Born approximation breaks down, we can obtain approximate (and strictly conservative) bounds by not considering non-perturbative couplings (i.e.~by not placing any exclusion limits on ranges small enough that only exponentially-large values of $\alpha$ would be excluded).

We can straightforwardly translate Eq.~\eqref{eq:YukawaSigmaTRelevant} into a heating rate, as we did for Coulomb scattering in the main text \cite{CoulombHeatingRate1, CoulombHeatingRate2, CoulombHeatingRate3}:
\begin{align}
    \frac{dU_\chi}{dt} \approx\,& 2\sqrt{2\pi} \alpha^2 G^2 \frac{M_{*,\tot}M_{\chi,\tot}}{R_*^3}\mathcal{F} \frac{m_*\sigma_*^2-m_\chi\sigma_\chi^2}{(\sigma_*^2+\sigma_\chi^2)^{3/2}} \ln\pfrac{4m_\chi^2(\sigma_*^2+\sigma_\chi^2)\lambda^2}{1+2\xi m_\chi^2(\sigma_*^2+\sigma_\chi^2)\lambda^2}. \label{eq:YukawaHR}
\end{align}
with $\mathcal{F}$ the form factor defined by Eq. \eqref{eq:FormFactorDef}, $\xi$ defined by Eq. \eqref{eq:YukawaCutoff} with $v_\rel^2 = \sigma_*^2+\sigma_\chi^2$ and $b_{\rm max}=\min(\lambda, R_*)$, and all of the velocity dispersions evaluated at $R_*$. Note that this reproduces the Coulomb result, Eq. \eqref{eq:TotalHeatingRate}, if the argument of the log is set to $\Lambda_G^2$; see Eq. \eqref{eq:CoulombLog}. We therefore define for convenience
\begin{align}
    \Lambda_Y^2 &= \frac{4m_\chi^2(\sigma_*^2+\sigma_\chi^2)\lambda^2}{1+2\xi m_\chi^2(\sigma_*^2+\sigma_\chi^2)\lambda^2}, \label{eq:YukawaLog}
\end{align}
which captures all of the effects of the both the finite halo size and finite force range (except insofar as they affect the halo mass and dark matter velocity dispersion).

\acknowledgments

The authors acknowledge support by NSF Grant PHY-2310429, Simons Investigator Award No.~824870, DOE HEP QuantISED award \#100495, the Gordon and Betty Moore Foundation Grant GBMF7946, and the U.S.~Department of Energy (DOE), Office of Science, National Quantum Information Science Research Centers, Superconducting Quantum Materials and Systems Center (SQMS) under contract No.~DEAC02-07CH11359. ZB was supported by the National Science Foundation Graduate Research Fellowship under Grant No.~DGE-1656518, by the Dr. HaiPing and Jianmei Jin Fellowship from the Stanford Graduate Fellowship Program, and by the DOE QuantISED program through the theory consortium ``Intersections of QIS and Theoretical Particle Physics'' at Fermilab. This manuscript has been authored by Fermi Forward Discovery Group, LLC under Contract No. 89243024CSC000002 with the U.S. Department of Energy, Office of Science, Office of High Energy Physics. 

\bibliographystyle{JHEP}
\bibliography{main}

\end{document}